\documentclass[preprint,prb,showpacs,showkeys,titlepage]{revtex4}
\usepackage{amsfonts}
\usepackage{amsmath}
\usepackage{amssymb}
\usepackage{graphicx}
\usepackage{subfigure}

\begin{document}

\title{Theoretical assessment on the possibility of constraining point defect energetics by pseudo-phase transition pressures}
\author{Hua Y. Geng}
\affiliation{National Key Laboratory of Shock Wave and Detonation
Physics, Institute of Fluid Physics, CAEP,
P.O.Box 919-102 Mianyang, Sichuan 621900, People's Republic of China}

\author{Hong X. Song}
\affiliation{National Key Laboratory of Shock Wave and Detonation
Physics, Institute of Fluid Physics, CAEP,
P.O.Box 919-102 Mianyang, Sichuan 621900, People's Republic of China}




\author{Q. Wu}
\affiliation{National Key Laboratory of Shock Wave and Detonation
Physics, Institute of Fluid Physics, CAEP,
P.O.Box 919-102 Mianyang, Sichuan 621900, People's Republic of China}


\begin{abstract}
Making use of the energetics and equations of state of defective uranium
dioxide that calculated with first-principles method, we demonstrate a possibility of constraining
the formation energy of point defects by measuring
the transition pressures of the corresponding pseudo-phase of defects.
The mechanically stable range of fluorite structure of UO$_{2}$, which dictates the maximum possible pressure
of relevant pseudo-phase transitions, gives rise to defect formation energies that
span a wide band and overlap with the existing experimental estimates.
We reveal that the knowledge about pseudo-phase boundaries can not only provide
important information of energetics that is helpful for reducing the
scattering in current estimates, but also be valuable for guiding
theoretical assessments, even to validate or disprove a theory.
In order to take defect interactions into account and to extrapolate the
physical quantities at finite stoichiometry deviations to that near the stoichiometry,
we develop a general formalism to describe the thermodynamics of a defective system.
We also show that it is possible to include interactions among defects in a
simple expression of point defect model (PDM) by
introducing an auxiliary constant mean-field.
This generalization of the simple PDM
leads to great versatility that allows one to study nonlinear effects of stoichiometry deviation
on materials' behavior.
It is a powerful tool to extract the defect energetics
from finite defect concentrations to the dilute limit.
Besides these, the full content of the theoretical formalism
and some relevant and interesting issues, including reentrant pseudo-transition,
multi-defect coexistence, charged defects, and possible consequence of
instantaneous defective response in
a quantum crystal, are explored and discussed.

\end{abstract}

\volumeyear{year}
\volumenumber{number}
\issuenumber{number}
\eid{identifier}
\maketitle

\section{INTRODUCTION}
\label{sec:intr}
Defects usually play a prominent role in various properties of a solid.
For this reason, the physics and chemistry of defects have been the subjects of much study for several decades.\cite{lopez88}
Many of these works focused mainly at the dilute limit, \emph{i.e.}, with a small defect concentration. This
is the case of interest in doped semiconductors and/or compounds in the immediate vicinity
of the stoichiometry. At certain conditions (usually with high temperature)
deviation from the stoichiometry can span over a wide range of chemical composition. Binary oxides CeO$_{2}$ and
UO$_{2}$ represent the paradigms for such kind of non-stoichiometry in the fluorite-related structures,
and many others exist as well.\cite{sorensen81} For a comprehensive understanding of these materials, knowledge
from just near the stoichiometry is insufficient. This is because many physical quantities depend
strongly on non-stoichiometry, and exhibit quite different behavior when at finite deviations.
On the other hand, defects at high pressure have received little attention so far, and our knowledge
about their general behavior at highly compressional conditions is very limited, in spite of
the fact that they are crucial for Earth modeling and for planetary evolution description,
where plenty of defects presenting in variety of non-stoichiometric
minerals in the interior of these celestial bodies.
The capability to capture correctly the energetics and other physical properties
across the whole stoichiometry range at different temperature-pressure conditions is an essential requirement for the
purpose of predicting and controlling the behavior of these complex materials.
Nevertheless, a general theoretical method for this purpose is still elusive.

Even at ambient conditions and near the stoichiometry, our understanding about nuclear oxides such as UO$_{2}$ and (U,Pu)O$_{2}$
is also limited and unsatisfactory.\cite{sorensen81}
Although there are many papers and reports have been published on various aspects of diffusion
in these oxides, a reasonable level of understanding has been reached only in the case of
oxygen, from which the formation energy of oxygen Frenkel pair can be deduced when the
migration energy of the corresponding diffusion process is known.\cite{sorensen81,matzke87,murch87,stratton87,tetot85}
Nevertheless, measuring oxygen diffusion in stoichiometric oxides is difficult because of the need to
maintain the stoichiometry, which is almost impossible for a large temperature range.\cite{matzke87}
The difficulty also lies in the interpretation of available experimental data. While
theoretical calculations can be applied to individual processes, transport and other data
often correspond to a superposition of several entangled processes, and makes
extraction of the desired information complicated.\cite{murch87,stratton87,tetot85}
In addition to chemical diffusion and self-diffusion, electrical conductivity\cite{stratton87} and neutron
scattering\cite{clause84} have also been performed to measure the defect concentrations and formation energies.
In all of these experiments, a poor understanding of the experimental conditions,
as well as the inherent difficulties in the measurement and subsequent interpretation of data,
have caused dispersed results.\cite{matzke87,tetot85,crocombette01}

There are also very few experimental data existed for cation defects in nuclear oxides.\cite{matzke87} For example in UO$_{2}$,
only crude estimates of the activation energy for uranium self-diffusion and migration energy of uranium vacancy
are available. With the aid of point defect model (PDM), one can extract the Schottky defect formation
energy from these estimates. But since a great uncertainty remains in the
experimental data, the reliability of the derived value is doubtful.\cite{crocombette01}

Progresses in density functional theory of electrons and in computational algorithms make it possible to
calculate the relevant energetics directly from quantum mechanics. Such kind of first principles
methods have provided better data than previous semi-empirical interatomic potentials,
and are comparable to experimental measurements. The electronic structures,\cite{petit10,dudarev98,prodan0607}
structural phase transformations\cite{geng07,song12}
and equations of state (EOS),\cite{geng07,geng11,geng12}
oxygen diffusion,\cite{gupta10,andersson09b,andersson12} and some defect clustering structures\cite{andersson12,andersson09,geng08,geng08b,geng08c}
have been modeled. Unfortunately, this advance never reached
a satisfactory level for defect energetics, even though a lot of efforts have been devoted to it.\cite{crocombette01,andersson09,geng08,freyss05,iwasawa06,dorado10,yu09,gupta07,brillant08,andersson11,nerikar09a,crocombette11,crocombette12}
It even cannot reproduce the experimental fact that oxygen defects
dominate the whole stoichiometry range.\cite{crocombette01,geng08}
Also one should note that including oxygen clusters can give rise to the expected predominance of oxygen
defects,\cite{geng08b,geng08c,geng11} but its relevance to the experimentally measured defect formation energy is unclear.\cite{crocombette12}
This discrepancy among experimental and theoretical results might be due to the limited accuracy of the theoretical assessments,
for example the possible meta-stable electronic states that could be encountered in calculations of strongly correlated
materials,\cite{larson07,jomard08,dorado09,geng2010}
the approximation employed to treat the partially localized 5\emph{f} orbitals,\cite{anisimov91,liechtenstein95,dudarev98b,zhou09,zhou11}
the variation of the charge state of defects,\cite{andersson11,nerikar09a,crocombette11,crocombette12}
and the small size of the simulation cell for defective structure modeling that currently accessible, \emph{etc.}
All of these might
render uncertain error in the final results.
But it also can arise from the error lying in the experimental estimates
that widely used as the benchmark for theoretical modeling: the data are scattering and not fully self-consistent,
and in some cases these estimated data cannot reproduce the fact of the predominance of
oxygen defects too.\cite{matzke87,tetot85,crocombette01}
This makes the problem entangled and very hard to treat with.
Therefore any approach that can constrain the defect energetics and reduce its
uncertainty is of decisive help for solving the problem.

In this report, we investigate the possibility of a such kind of method by examining the physics
that governs the pressure-driven pseudo-transitions\cite{geng12} between different defect species. By establishing a
theoretical relationship of the pseudo-phase boundary on defect formation energies, we show that
strict constraints can be imposed on these energy parameters, which might then be used to refine the
experimental estimates.
To achieve the final goal, however, one has first to measure the curve of pseudo-transition pressures,
and obtain accurate equations of state of defects, then by making use of the theoretical method
that we will present below, to get reliable estimate of defect formation energies. In this sense this report
is the first step---but also the most important step---towards this accomplishment. By
developing the theoretical basis of this constraining procedure, it not only provides us a new angle to
understand the long-standing problem, but also establishes a general method in treating highly
defective materials under high pressures. For the clarity of discussion, we will first present
a simple theoretical framework using PDM that is intuitive and easy to understand,
then a generalization to the general case will be made.

In Sec.\ref{sec:simple_model} we discuss the PDM
that allows us to calculate the pseudo-phase diagram of point defects, and then establish a relationship between
pseudo-phase boundary and defect energetics. Using this powerful tool, the influence
of intrinsic defect formation energy on pseudo-transition is then investigated in Sec.\ref{sec:PDM}.
The obtained information represents a constraint on the possible value of the formation energy
of the defects. Though our
discussions are mainly focused on uranium dioxide in this paper, we also extend the investigation by
considering virtual models to explore other
interesting phenomena such as reentrant pseudo-transition and multi-defect coexistence
in Sec.\ref{sec:virtual system}. It is well known that PDM does not take defect-defect interactions
into account and can be applied to only the vicinity of the stoichiometry.
In order to deal with highly defective region and extrapolate the energetic and thermodynamic information obtained at finite stoichiometry deviations
to the dilute limit, a general formalism that can treat defect interactions is developed in Sec.\ref{sec:defect interactions}.
This generalization is necessary for a realistic description of the nonlinear dependence of thermodynamic properties on non-stoichiometry.
A brief discussion and remarks on charge state of defects, as well as other relevant issues, are then given in Sec.\ref{sec:discussion},
which is followed by a summary of the main conclusions.

\section{Simple THEORETICAL FRAMEWORK}
\label{sec:simple_model}
At dilute limit where defect concentration is negligibly small, imperfectness in
crystal has little impact on thermodynamic and mechanical properties, despite a profound modification
on electrical conductivity and/or magnetism might often occur. At large stoichiometry deviation,
the defect concentration is governed by the ratio of chemical compositions rather than by thermal excitation,
therefore the density of defects could be enormous. In this case, noticeable influence on general
thermodynamic quantities can be expected, so as on relevant mechanical properties.

To understand the general effects of non-stoichiometry on material's behavior, we require a physical model that expresses
the defect density as a function of external conditions---usually the hydrostatic pressure and
temperature, and how the presentation of defects modifies thermodynamic functions
such as enthalpy or Gibbs free energy. Having such a function that incorporated defect
effects, all relevant thermodynamic properties can be derived straightforwardly.
In this section we first present the basic
picture by considering the simple PDM.
A general formalism will be developed in Sec.\ref{sec:defect interactions}.

In PDM,\cite{matzke87,lidiard66} the spatial size of an individual defect is assumed to be of zero dimension, and all interactions among them are neglected.
In this simple model, defect concentrations are determined by the corresponding formation energy of isolated defects.
Considering a structure that contains one defect of type $i$, its Gibbs free energy can be written as
\begin{equation}
  G_{i}(P,T)=E_{c}(V)+F_{ph}(V,T)+PV,
\end{equation}
in which $P$, $T$, and $V$ stand for hydrostatic pressure, temperature, and volume of the simulation
cell, respectively. The cohesive energy at zero Kelvin in static approximation reads
\begin{equation}
  E_{c}(V)=-D+\frac{9}{8}B_{0}V_{0}\left[\left(\frac{V_{0}}{V}\right)^{\frac{2}{3}}-1\right]^{2}
  \label{eq:coldE}
\end{equation}
when expressed in Birch-Murnaghan equation (other EOS model can be used as well). Here variables with subscript 0 denote the corresponding
value in the equilibrium condition of zero pressure, and $B$ is the bulk modulus.
The contribution of lattice dynamics to the free energy can be approximated in the Debye model as
\begin{equation}
  F_{ph}(V,T)=3k_{B}T\ln\left[1-\exp(-\Theta_{D}/T)\right]-k_{B}Tf\left(\Theta_{D}/T\right)+\frac{9}{8}k_{B}\Theta_{D},
\end{equation}
where $k_B$ is the Boltzmann constant and $f$ the Debye function. The Debye temperature can be
evaluated approximately by\cite{geng05,moruzzi88}
\begin{equation}
  \Theta_{D}=\Theta_{D}^{p}\left[\frac{BM^{p}}{B^{p}M}\left(\frac{v}{v^{p}}\right)^{1/3}\right]^{1/2}.
\end{equation}
Here the superscript $p$ denotes the reference state (here the defect-free UO$_2$),
$v$ is the effective volume per atom,
and $M$ is the effective atomic weight. The parameters in these equations, namely $D$, $B_{0}$, and $V_{0}$,
can be obtained by fitting to \emph{ab initio} results of density functional theory, while $\Theta_{D}^{p}$
can be taken from x-ray diffraction measurement.\cite{serizawa99,serizawa00} The details of determining the value
of these parameters have been discussed and presented in Ref.\onlinecite{geng11}.

Having known $G_{i}$, the formation Gibbs free energy (FGE) of intrinsic point defects can be
constructed. For example, the FGE of a Frenkel pair (FP) of species $X$ is then expressed as
\begin{equation}
 \Delta G_{\mathrm{X{\_}FP}}=G^{N-1}_{X_{v}}+G^{N+1}_{X_{i}}-2G^{N},
 \label{eq:GFP}
\end{equation}
and for the Schottky defect ({S}) as (taking UO$_{2}$ as the example)
\begin{equation}
 \Delta G_{\mathrm{S}}=G^{N-1}_{\mathrm{U}_{v}}+2G^{N-1}_{\mathrm{O}_{v}}-3\frac{N-1}{N}G^{N}.
 \label{eq:Gs}
\end{equation}
Here $N$ denotes the number of atoms in a defect-free cell
and $G^{N}$ is the corresponding Gibbs free energy, $G^{N\pm 1}_{X_{v},{X_{i}}}$
is the Gibbs free energy of the cell containing the respective defect.
In a closed regime where no particle exchange with the exterior can occur,
the defect concentration must satisfy\cite{crocombette01,geng11,geng08}
\begin{align}
  & [V_{O}][I_{O}]=\exp\left(\frac{-\Delta G_{\mathrm{O{\_}FP}}}{k_{B} T}\right)\label{eq:PDM1},\\
  & [V_{U}][I_{U}]=\exp\left(\frac{-\Delta G_{\mathrm{U{\_}FP}}}{k_{B} T}\right)\label{eq:PDM2},\\
  & [V_{O}]^{2}[V_{U}]=\exp\left(\frac{-\Delta G_{\mathrm{S}}}{k_{B} T}\right)\label{eq:PDM3}.
\end{align}
The composition equation that expressed in point defect populations is
\begin{equation}
x=\frac{2\left([V_{U}]-[I_{U}]\right)+[I_{O}]-2[V_{O}]}{1-[V_{U}]+[I_{U}]},
\label{eq:PDMx}
\end{equation}
where $x$ is the stoichiometry deviation (for example that in UO$_{2+x}$). Notice that Eq.(\ref{eq:PDMx}) is different
from the conventional definition of
\begin{equation}
  x=2\left([V_{U}]-[I_{U}]\right)+[I_{O}]-2[V_{O}],
  \label{eq:PDM4}
\end{equation}
which is valid only when no cation defect is involved. We thus complete the formalism of PDM, in which the defect concentrations are
determined by solving Eqs.(\ref{eq:PDM1}$\sim$\ref{eq:PDMx}).

Since no interaction among defects has been taken into account in PDM, the total Gibbs free energy
of a defective system is a linear superposition
of the contribution of each individual defect. That is,
\begin{equation}
  G\approx G_{0}+\sum_{i}\frac{\Delta G_{i}}{n_{i}^{\mathrm{ref}}}n_{i},
  \label{eq:Gtot}
\end{equation}
where $G_{0}$ is the Gibbs free energy of the defect-free matrix.
The defect concentration $n_{i}$ runs over $[I_{O}]$, $[V_{O}]$, $[I_{U}]$, and $[V_{U}]$, respectively,
with the superscript ``$\mathrm{ref}$'' indicates the corresponding value in
a defective reference system,
and $\Delta G_{i}=G_{i}^{\mathrm{ref}}-G_{0}$. From Eq.(\ref{eq:Gtot})
thermodynamic quantities as a function of stoichiometry deviation $x$ can be derived.

\section{pseudo-phase diagram and pseudo-transition}
\label{sec:PDM}
In defective crystals, distinction of the physics mainly originates from the predominant defect species.
Thanks to the exponential dependence of defect concentrations on the formation energy, most regions
in the phase space spanned by temperature, pressure, and chemical composition ($T$-$P$-$x$) are dominated
by only one type of defect. One can then use the concept of pseudo phase to simplify
the description of defective (non-stoichiometric) materials.\cite{geng12} In this picture,
each pseudo phase corresponds to a region that is
governed by a homogeneous distribution of a \emph{single} type of defect.
Here no effects of migration and creation or annihilation of defects are considered, which is
justified if we focus mainly on the \emph{long-time averaged} properties only.

With variation of the thermodynamic conditions of $T$, $P$, and $x$, the predominant
defect might change from one type into another, \emph{i.e.}, pseudo phase transition (PPT) might take place.
Physical quantities that affected by defects also change rapidly along this transition.\cite{geng12}
From this perspective, the physics of defects can be greatly simplified to that of
each individual pseudo-phases and their respective behavior at the PPT.
It is thus important to understand the extent of the control region of each pseudo phase, namely,
the pseudo phase diagram (PPD).
From Eqs.(\ref{eq:PDM1}$\sim$\ref{eq:PDMx}), it is evident that such a diagram is completely described by
the energetics of each defect. Conversely, if we know the PPD, then constraints
on defect formation energies can be established.

It is natural to define the transition point of a PPT at where defect
concentration increasing/decreasing to the half of its
saturate value, then
the corresponding pseudo-phase boundary (for example in UO$_{2+x}$) is determined by the following equation
\begin{equation}
  \frac{1}{2x}=\exp\left( \frac{-\Delta G_{\mathrm{S}}+2\Delta G_{\mathrm{O{\_}FP}}}{k_{B}T}\right)
\label{eq:ppbxp}
\end{equation}
for a transition between $\mathrm{U}_{v}$ and $\mathrm{O}_{i}$ when $x>0$, and
\begin{equation}
  -\frac{2}{x}=\exp\left( \frac{\Delta G_{\mathrm{S}}-\Delta G_{\mathrm{U{\_}FP}}}{k_{B}T}\right)
\label{eq:ppbxn}
\end{equation}
between $\mathrm{U}_{i}$ and $\mathrm{O}_{v}$ when $x<0$ (only point defects are considered).
Because all $\Delta G$ are functions of $T$ and $P$ (and also depend on $x$ via defect interactions,
which we discarded here but will discuss in detail below), solutions of
Eq.(\ref{eq:ppbxp}) and (\ref{eq:ppbxn}) provide a set of constraints on formation energy of intrinsic defects.

\subsection{Realistic system: UO$_{2}$}

\begin{figure}
  \includegraphics*[width=3.5 in]{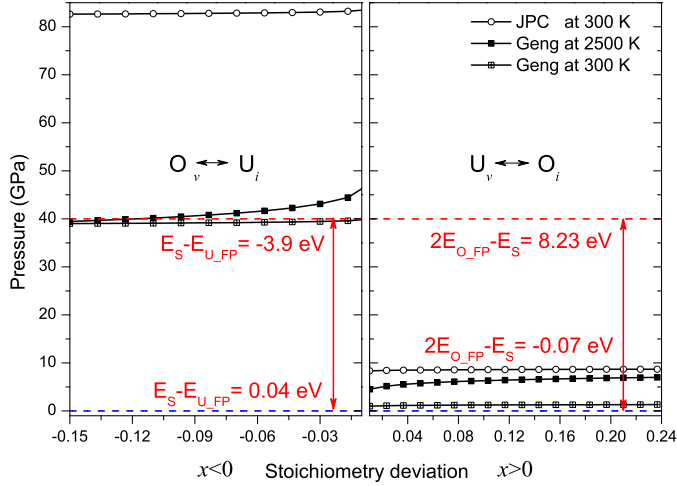}
  \caption{(Color online) Comparison of the pseudo-phase diagram of UO$_{2+x}$ calculated with Crocombette (JPC)\cite{crocombette12} and Geng's\cite{geng11} formation
  energies of defects. The constraints on intrinsic defect formation energies when the pseudo transition
  pressure bounded between 0 and 40\,GPa are also shown.}
  \label{fig:ppd}
\end{figure}

This subsection is devoted to a realistic system of UO$_{2}$, where involved parameters are
obtained by density functional theory calculations. Since early experiments
were driven by application of UO$_{2}$ as a nuclear fuel, most investigations focused mainly on
formation energy of intrinsic defects, which closely relates to the
parameter $D$ in Eq.(\ref{eq:coldE}). To be consistent with conventional notations,
we use $E_{\mathrm{S}}$, $E_{\mathrm{O{\_}FP}}$, and $E_{\mathrm{U{\_}FP}}$
to denote the formation energy of Schottky defect, oxygen Frenkel pair, and uranium Frenkel pair at zero pressure and temperature, respectively.
Based on the defect energetics calculated by Geng \emph{et al.} for uranium dioxide,\cite{geng11}
the PPD on the $P$-$x$
plane are evaluated and shown in Fig.\ref{fig:ppd}.
It can be seen that in the region of $x<0$ an increasing of temperature from 300 to 2500\,K has little
influence on the pseudo boundary. However, in the $x>0$ side, such a size of change in $T$
leads to an increase of the pseudo transition pressure about 5\,GPa.
Except that, the impact of $x$ on PPD at the PDM level of approximation is small, and with an opposite
trend for the hypo- and hyper-stoichiometry region.

There are a few other theoretical assessments of the defect energetics for UO$_{2}$ available
in literature.\cite{crocombette01,andersson09,geng08,freyss05,iwasawa06,dorado10,yu09,gupta07,brillant08,andersson11,nerikar09a,crocombette11,crocombette12}
Unfortunately in most cases only formation energy at 0\,GPa and 0\,K were given, from which,
however, one cannot determine the pseudo transition pressure because the information about the variation
with temperature and pressure is lost. Under an assumption that the compression behavior and
phonon contribution are the same for all of these calculations, which is a simple but reasonable
approximation, we can estimate the corresponding pseudo transition boundary by
adjusting the $D$ in Eq.(\ref{eq:coldE}) accordingly to yield the respective formation energy at 0\,GPa
and 0\,K using Geng's equation of state.\cite{geng11}
Here we choose the data of J. P. Crocombette (JPC) for the purpose of comparison, since it is a typical one that has considered
possible charge states of defects, thus produced a formation energy of oxygen Frenkel pair and
Schottky defect that seems in a good agreement with experimental estimates.\cite{crocombette12}
By adjusting $D$ to
reproduce the $E_{\mathrm{S}}$ and $E_{\mathrm{O{\_}FP}}$ of JPC, we obtained
the estimated PPD of JPC's data. Note here we have made an assumption that the charge state of
each defect is fixed during compression or heating, and the formation energy of uranium interstitial
was taken from Geng's data because in JPC's work no value for this defect type was given.

The PPD calculated with JPC's formation energy is drawn in Fig.\ref{fig:ppd} for comparison. At $x<0$ side, the transition
from $\mathrm{O}_{v}$ to $\mathrm{U}_{i}$ takes place at a much higher pressure. This is reasonable
since JPC's data has a lower formation energy for $\mathrm{O}_{v}$ than Geng's evaluation, which gives rise to a stronger
stability of this defect. At the $x>0$ side, uranium vacancy was predicted to be the predominant
defect at low pressure, and the transition to $\mathrm{O}_{i}$ occurs at about 8\,GPa. This
result is consistent with previous PDM evaluations at zero hydrostatic pressure, where
$\mathrm{U}_{v}$ was predicted to be the major defect component. Nevertheless, prevailing $\mathrm{U}_{v}$
is contradictory to
the experimental observation that oxygen defect should dominate this region, indicating that
the experimental estimate of the defect formation energy might be inconsistent in itself.\cite{matzke87,crocombette01,tetot85}
This difficulty could stem from the procedure of extraction defect energetics from diffusion
measurements. Usually the employed physical models were very simple and might lead to inaccurate explanation of the
measured data.\cite{matzke87,murch87,stratton87,tetot85}

On the other hand, as mentioned above, PPD provides valuable information about possible
range of the defect formation energy. For UO$_{2}$, the PPT in fluorite structure (if exists)
should be bounded between a pressure range of 0 and 40\,GPa, because at higher pressures UO$_{2}$
transforms into $Pnma$ phase,\cite{idiri04} which is followed by an iso-structural transition,\cite{geng07} and finally
converges to isotropic $P6{_3}/mmc$ structure\cite{song12} according to recent theoretical predictions.
Under this restriction, the formation energy of intrinsic defects
should satisfy
\begin{equation}
  -3.9\,\mathrm{eV} \leq E_{\mathrm{S}}-E_{\mathrm{U{\_}FP}}\leq 0.04\,\mathrm{eV}
  \label{eq:ppd1}
\end{equation}
when $x<0$, and
\begin{equation}
  -0.07\,\mathrm{eV} \leq 2E_{\mathrm{O{\_}FP}} -E_{\mathrm{S}}\leq 8.23\,\mathrm{eV}
  \label{eq:ppd2}
\end{equation}
when $x>0$ for a PPT to occur between a pressure of 0 and 40\,GPa. These constraints are marked in Fig.\ref{fig:ppd} together with dashed lines
that indicate the corresponding bounded range of pressure.

\begin{figure}
  \includegraphics*[width=3.0 in]{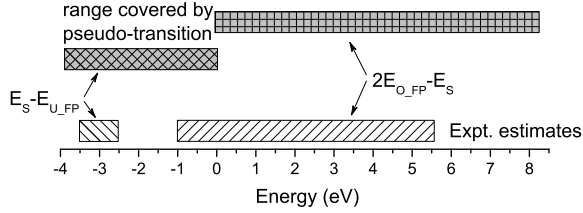}
  \caption{A schematic diagram which illustrates the range of defect formation energy that is covered
  by pressure-driven pseudo-transition (upper part) and the corresponding experimental estimates (lower part) in non-stoichiometric UO$_{2}$.
  }
  \label{fig:TP}
\end{figure}

Theoretical assessment explicitly suggests that $\mathrm{U}_{v}$ definitely becomes unfavorable under compressional conditions.
Therefore the
experimental observation that $\mathrm{O}_{i}$ prevails at $x>0$ region
implies that there should be no PPT from $\mathrm{U}_{v}$ to $\mathrm{O}_{i}$ at any pressures greater than zero.
Then from Eq.(\ref{eq:ppd2}) one gets
\begin{equation}
  E_{\mathrm{S}} \geq 2E_{\mathrm{O{\_}FP}} +0.07\,\mathrm{eV},
  \label{eq:ppd3}
\end{equation}
which puts a strong constraint on possible value of defect formation energies.
For example the experimental estimates of $E_{\mathrm{O{\_}FP}}$ lie in between
$3.0\sim 4.6$\,eV and $E_{\mathrm{S}}$ between $6.0\sim 7.0$\,eV.\cite{matzke87,clause84}
If we take $E_{\mathrm{S}}$ as 7.0\,eV, then $E_{\mathrm{O{\_}FP}}$ must be less than 3.5\,eV.
This value is, however, incompatible with the most reliable experimental assessments.\cite{stratton87,murch87,clause84} On the other hand,
if we take the neutron scattering measurement\cite{clause84} of 4.6\,eV as a reliable estimate for $E_{\mathrm{O{\_}FP}}$, then $E_{\mathrm{S}}$
must be greater than 9.3\,eV. This in turn disqualifies most theoretical estimates with charged defects.\cite{andersson11,crocombette12}
In a word, all of these indicate that we need further scrutiny on these estimates,
and any alternative and/or complemental information on defect energetics are decisive
to reach the final conclusion.
Inequalities of Eqs.(\ref{eq:ppd1}$\sim$\ref{eq:ppd2}) cover most range of the experimental estimates, as shown
in Fig.\ref{fig:TP}. Thus it can provide new understanding about this issue if we can
measure the compression-driven PPT of point defects experimentally.

\subsection{Virtual system: model study}
\label{sec:virtual system}

Defective behavior of materials at high pressure is determined by the variation of
defect formation enthalpy with compression. It is also affected by possible structural transitions
of the matrix. Above discussion elaborated what might happen in a compressed non-stoichiometric UO$_{2}$.
In other materials, however, much more complex phenomena can be expected. From a theoretical perspective,
it is helpful to explore all possibilities in order to grasp the general feature
of the physics of defects. In the simple PDM approximation, the physics is mainly determined by the parameters
appeared in Eq.(\ref{eq:coldE}), \emph{i.e.}, the value of $D$, $B_{0}$, and $V_{0}$
of each defective configurations that were employed to derive the formation Gibbs free energy (FGE). Therefore we can arbitrarily
alter these parameters to probe other interesting behaviors of defects that are allowed
in theory but not in the realistic UO$_{2}$.

\subsubsection{Reentrant transition}

The first phenomenon we would like to discuss about is reentrant transition. It is a rare type
of phase change even for a conventional physical state, where one phase that has been transformed
into another re-appears.
On the phase diagram the corresponding phase boundary is a reentrant curve.
Analogous phenomenon
can also take place in PPT, where the predominant defect species
firstly changes into another type, and then transforms back.
The condition of this transition
is completely governed by Eqs.(\ref{eq:PDM1}$\sim$\ref{eq:PDM3}).
Put it explicitly, if the derived equation [Eq.(\ref{eq:ppbxp}) or (\ref{eq:ppbxn})]
has multiple solutions, then the corresponding PPT is reentrant.

\begin{figure}
  \includegraphics*[width=3.0 in]{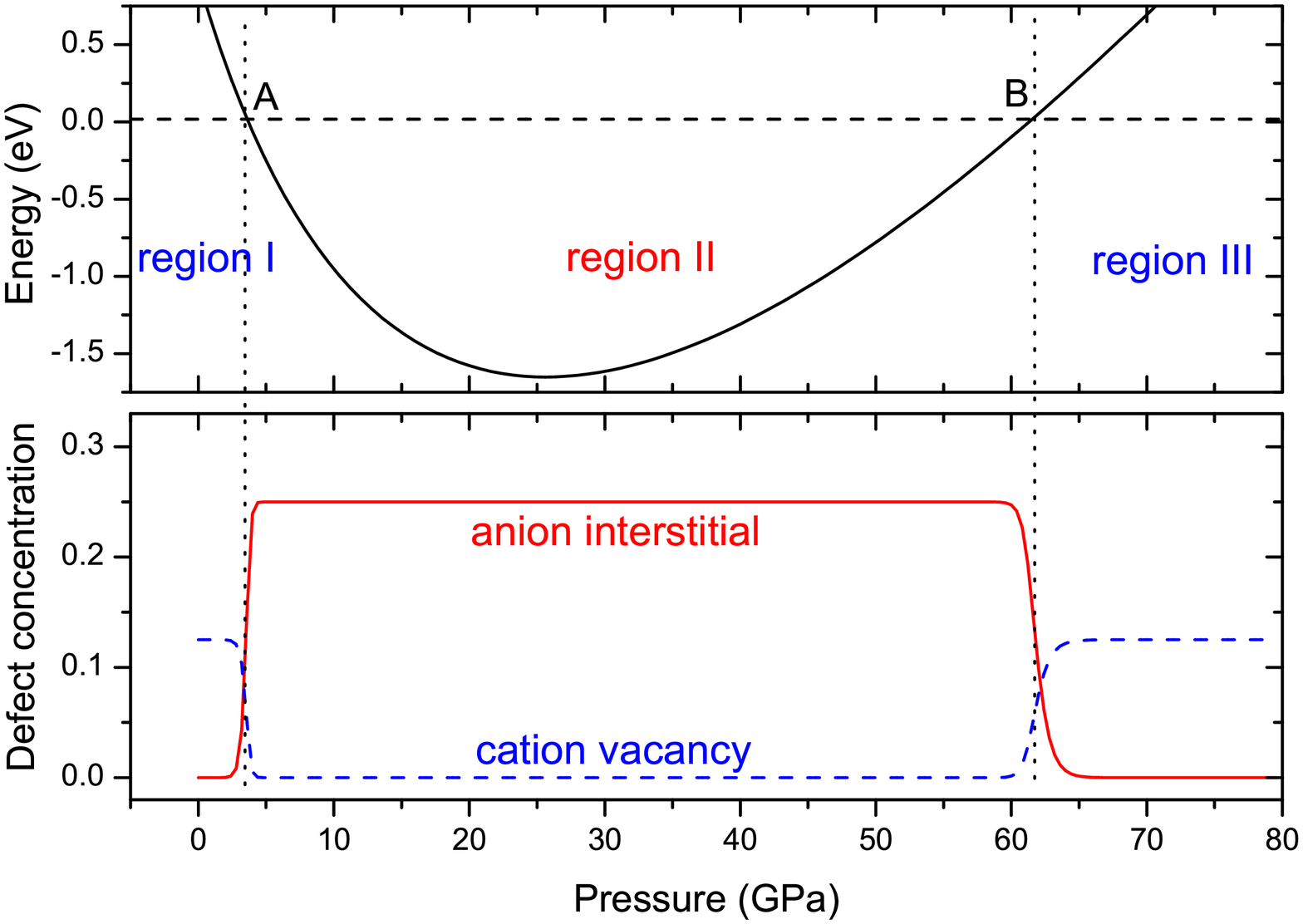}
  \caption{(Color online) Pressure-driven reentrant pseudo-transition of defects at 300\,K and $x=0.25$. Upper panel: energy
  variation; lower panel: defect concentrations.
  }
  \label{fig:reentry}
\end{figure}

Figure \ref{fig:reentry} demonstrates a virtual dioxide compound that has
a reentrant PPT. In the upper panel,
a graphical solution of Eq.(\ref{eq:ppbxp}) is drawn where the solid line
is the term of $-\Delta G_{\mathrm{S}}+2\Delta G_{\mathrm{O{\_}FP}}$. Another term of
$k_{B}T\ln({\frac{1}{2x}})$ is also shown as a dashed line in the figure.
The points of intersection A and B correspond to the solutions of Eq.(\ref{eq:ppbxp}),
which also are the locations
where PPT taking place.
In the lower panel, the change of defect concentrations along compression is illustrated,
from which one can clearly see that cation vacancy re-appears at higher pressures.

\begin{figure}
  \includegraphics*[width=3.0 in]{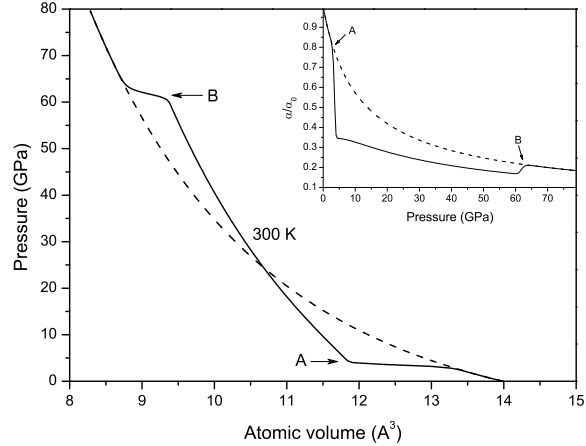}
  \caption{Isothermal compression curve of a reentrant PPT as showing in Fig.\ref{fig:reentry}.
  Inset: the relative variation of thermal expansivity.
  }
  \label{fig:PV_ppt}
\end{figure}

It is necessary to point out that this result
was obtained by subtracting 145\,GPa from the bulk modulus of all defective UO$_{2}$ configurations,
thus might be an artifact. Nevertheless, such a virtual model can help us acquire a profound
understanding of material's behavior that having reentrant PPT, if it exists.
The resultant modifications on EOS and thermodynamic properties across this transition region are interesting.
Figure \ref{fig:PV_ppt} illustrates the compression behavior of the same PPT
as in Fig.\ref{fig:reentry}, in which the restoring of the compression curve at high pressure end is evident.
There are two volume-collapses in the reentrant PPT at point A
and B, respectively. Between A and B, the curve
is steeper. This is consistent with the requirement for a pressure-driven reentrant transition
to occur, namely, the intermediate phase should have a larger bulk modulus and a smaller equilibrium volume.
This condition guarantees a double volume-collapse
along compression, which is a necessity for a first-order reentrant transition. Other
thermodynamic properties also show discontinuous or quasi-discontinuous jump at the PPT points.
The inset in Fig.\ref{fig:PV_ppt} draws relative variation of thermal expansivity $\alpha$
with respect to its initial value along the 300\,K
compression curve. It deviates from the trend of the initial phase (as the dashed line shows) at point A and plunge
to a new value, but at point B it jumps back to the previous curve, a key feature
of reentrant transition. Other physical quantities, such as specific heat and compressibility,
demonstrate similar characteristics.

\subsubsection{Coexistence of defects}

The picture of pseudo-phase of defects is only valid when the dominating region of the associated
defect type is well defined. For point defects at low temperature,
it is usually the case. However, with elevated temperature and/or when complex defect clustering is involved,
competition might lead to coexistence of different defect species, where the
notation of pseudo phase could lose its physical importance.

For realistic UO$_{2}$, pseudo-phase can always be defined, whether oxygen clusters (\emph{e.g.}, COT-o
cluster) involved or not. At high temperatures, the zone of pseudo phase boundary becomes wide,
and renders the PPT as a smooth crossover.\cite{geng12} In spite of this, the material behavior still can
be understood within the framework of pseudo phase.
In some conditions, however, a situation
that one defect species appears but never gains the dominant role might be possible. This
will completely invalidate the picture of pseudo phase,
and a detailed analysis of defect concentrations becomes necessary (in contrast to this, the
defect concentrations are determined by stoichiometry deviation if pseudo phase can be applied).

\begin{figure}
  \includegraphics*[width=3.0 in]{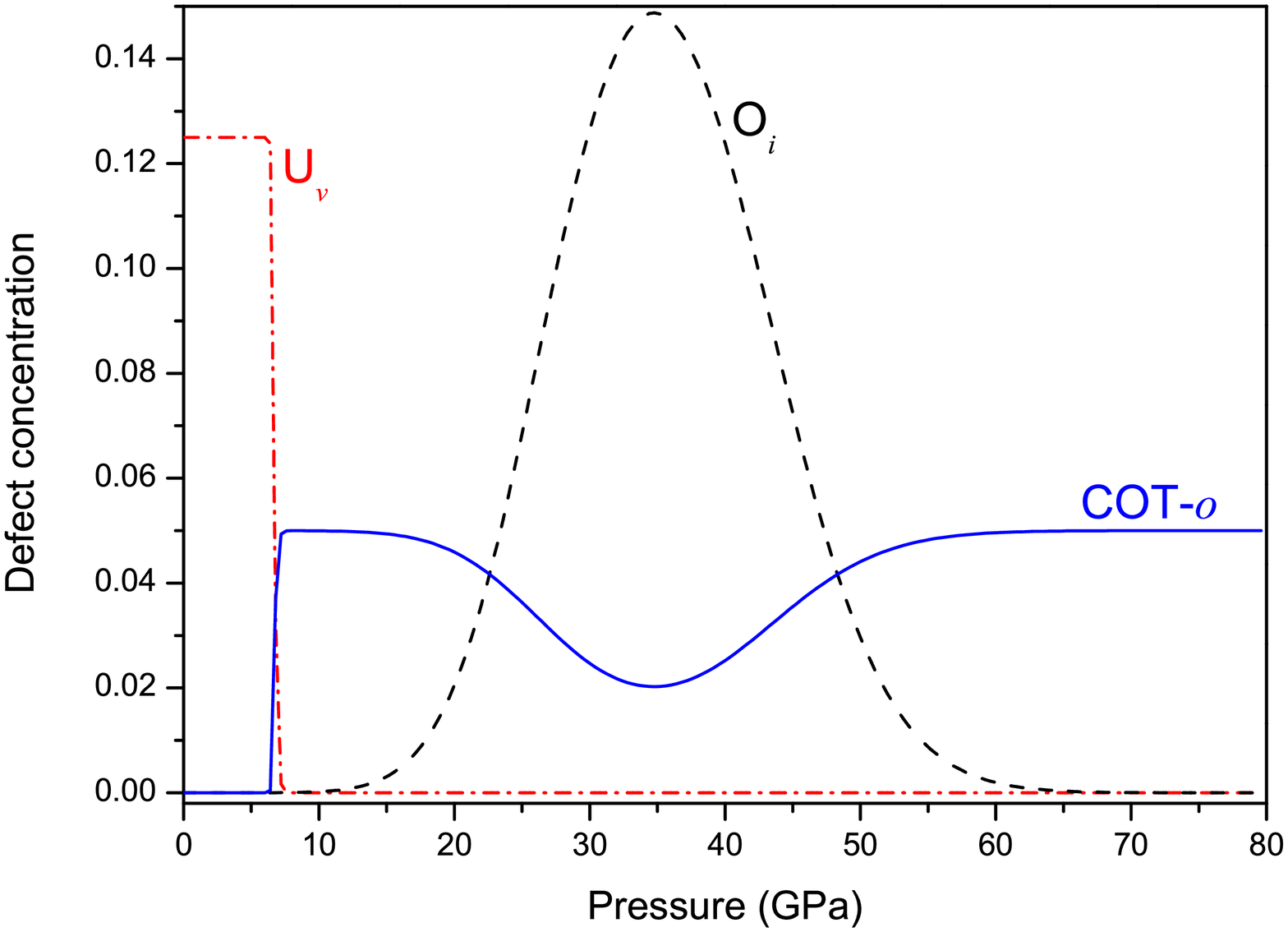}
  \caption{(Color online) Multi-defect coexistence at 300\,K and $x=0.25$, see text for details.
  }
  \label{fig:coext}
\end{figure}

An example of this is presented in Fig.\ref{fig:coext}, which was generated with JPC's formation energy data.\cite{crocombette12}
In addition to that, the value of COT-o cluster\cite{geng11} was also reduced by 4.865\,eV, namely, we have artificially
decreased the stability of oxygen clusters. The result is that oxygen interstitial O$_{i}$ gets some promotion, but not
enough to become the major one, and persists only within a narrow pressure range. It is evident from
Fig.\ref{fig:coext} that in this virtual system U$_{i}$ and COT-o (at least for most
pressure range) can be depicted using pseudo phase, and there is a PPT from U$_{i}$ to COT-o
at a pressure of 7\,GPa. On the other hand, O$_{i}$, which appears from 20 to 50\,GPa,
depletes the concentration of COT-o slightly. But it never gains the dominance, and the phenomenon must be taken as a defect coexistence
rather than being a PPT. To understand the difference between these is crucial for a correct
description of defective materials using the concept of pseudo phase.

\section{defect interactions}
\label{sec:defect interactions}

Above discussion is based on the approximation of PDM, which is valid at the dilute limit.
With an increase of defect concentrations, however, interaction between defects
becomes significant. This could lead to a severe deviation from the prediction of PDM.
On the other hand, non-stoichiometry might affect the EOS and energetics of defects, thus modify
the PPD. To obtain the correct defect behavior near the stoichiometry,
a close cooperation between experimental measurements and theoretical analysis is necessary. For
that purpose, a thorough and comprehensive understanding of defect interaction is crucial.
This is because with such an accurate information, we can deduce the formation energy
at dilute limit from the measured PPD at finite stoichiometry deviations. It is worthy of
mentioning that the same extraction procedure can be done using the simple PDM, but the
resultant error is usually quite large.
In this part of the paper, we will first derive a general formalism that includes
defect interactions, then we will show that the conventional PDM can be adapted to take
these interactions into account effectively with a simple constant mean-field approximation,
which greatly expands the applicability of PDM.

\subsection{General formalism}

If one is interested mainly in macroscopic properties of a defective material that averaged over a long enough time scale,
the most important contribution comes from defects in static distributions. For a classical crystal
at finite temperature, defects have nonzero probability of migrating between available sites.
This alters the dynamics and transport property of the system. Nevertheless, diffusion
does not modify the overall macroscopic properties very much as long as the distribution of
defects is still in an equilibrium state. This is because migration is a transient process and in most time defects are trapped in their equilibrium sites.
In a migration process, defects induce dynamic deformations in the local lattice, and scatter
with phonon. This effect might drive phonon away from its equilibrium distribution.
However, considering that this change is very small compared to thermal fluctuations, and our
interested time scale is many orders longer than the relaxation time of phonon, it is
justified to ignore this effect.

The main assumptions in the theory include: (a) we work on a lattice model, (b) only static distribution of defects is considered,
(c) the distribution
must be homogeneous, (d) dynamic effects due to defect migration are ignored, (e) no defect creation and annihilation
are considered, (f) when defect clusters are involved, taking them as single objects,
namely, ignore association and/or dissociation effects.

With these assumptions, let us consider a grand canonical ensemble on a lattice
in which the number of particles fluctuating around the
average value. Assume there are in total $K$ sublattices, which are occupied by $M$ species of particles.
Using $N_{i}^{m}$ denotes the number of $m$-th kind of particle that siting on the $i$-th sublattice,
where $i$ can take any value from 1 to $K$ and $m$ varies from 1 to $M$. It is evident that each system in the ensemble
is characterized by the occupation of particles on the lattice, and thus the ensemble
can be grouped according to $\{N_{i}^{m}\}$. That is to say, each system can be
labeled uniquely by a set of $\{N_{i}^{m}\}$ together with an auxiliary index $s$, where $s$ runs over all possible
configurations that has the same $\{N_{i}^{m}\}$.
In this way, if using $\rho_{q}$ denotes
the probability of configuration $\left(\{N_{i}^{m}\},s\right)$ in the ensemble,
where  $\left(\{N_{i}^{m}\},s\right)$ has been shortened as $q$ for brevity,
then the Gibbs free energy of the ensemble is
\begin{equation}
  G=\sum_{q} \left[\rho_{q} F_{q} +k_{B}T\rho_{q}\ln\left(\rho_{q}\right)\right].
  \label{eq:gibbs1}
\end{equation}
Here $F$ is the Gibbs free energy of an individual configuration, and the contribution
of configurational entropy has been separated out and presents as the second term in Eq.(\ref{eq:gibbs1}).
The thermodynamically equilibrium state is achieved when $G$ takes a minimum.
The normalization condition requires $\sum_{q} \rho_{q}=1$. If $N^{m}$ denotes the total
number of the $m$-th particle in the ensemble, then $N^{m}=\sum_{i}\sum_{q}N_{i}^{m}\rho_{q}$.
Introducing Lagrange multipliers $\mu_{m}$ and $\lambda$, the minimization equation of $G$ becomes
\begin{equation}
  \frac{\delta G}{\delta \rho_{q}} + \sum_{m} \mu_{m}\frac{\delta (N^{m}-\sum_{i}\sum_{q}N_{i}^{m}\rho_{q})}{\delta \rho_{q}}
  +\lambda\frac{\delta(1-\sum_{q} \rho_{q})}{\delta \rho_{q}}=0.
\end{equation}
Making use of Eq.(\ref{eq:gibbs1}), this leads to
\begin{equation}
  \rho_q = {\exp\left(\frac{-F_{q} + \sum_{i}\mu_{m}N_{i}^{m}}{k_{B}T}\right)}/{\Xi},
  \label{eq:rho}
\end{equation}
where the partition function $\Xi = \sum_{q} {\exp\left(\frac{-F_{q} + \sum_{i}\mu_{m}N_{i}^{m}}{k_{B}T}\right)}$.
It can be shown that $\mu_{m}$ is just the chemical potential of
the $m$-th type of particle. The free energy of the $q$-th
configuration, $F_{q}$, which consists of the cold crystal energy at 0\,K and phonon contributions to the
internal energy and entropy (as well as a term of $PV$, if at finite pressure),
can be calculated with modern first-principles methods. Since
defects in an individual configuration manifest as an imperfect occupation of the lattice sites,
the equilibrium thermodynamics of defects is therefore completely described by Eq.(\ref{eq:rho}),
in which the contribution of defect interactions arises from $F_{q}$.

When the energy scale of temperature is much smaller than the formation energy of defects, which
usually is the case for most applications, the partition function is dominated by the perfect occupation (\emph{i.e.}, the ground-state),
namely $\Xi \simeq {\exp\left(\frac{-F_{0} + \sum_{i}\mu_{m}{N_{0}}_{i}^{m}}{k_{B}T}\right)}$,
where the subscript 0 denotes the groundstate.
In this situation Eq.(\ref{eq:rho}) reduces to
\begin{equation}
  \rho_{q} = \exp\left(-\frac{\Delta G}{k_{B}T}\right),
  \label{eq:rhoq}
\end{equation}
where the FGE of configuration $q$ is given by $\Delta G=F_{q}-F_{0}+
\sum_{i}\mu_{m}(N_{i}^{m}-{N_{0}}_{i}^{m})=\Delta F +\sum_{i}\mu_{m}\Delta N_{i}^{m}$.
For single point defect $\Delta N =\pm 1$. Thus the FGE of intrinsic defects such as Frenkel pairs and Schottky
defect can be derived straightforwardly from Eq.(\ref{eq:rhoq}), and have the same form as Eqs.(\ref{eq:GFP}$\sim$\ref{eq:Gs}).
Furthermore, equations similar to PDM [Eqs.(\ref{eq:PDM1}$\sim$\ref{eq:PDM3})] also can be constructed easily using
Eq.(\ref{eq:rhoq}), indicating that the simple formalism of PDM is
much more flexible than what it was originally proposed.

It is necessary to point out that for a system defined on a lattice, there is an orthogonal and
complete basis set called correlation functions, which are an alternative but powerful representation of all possible
occupation configurations of particles on the lattice using increasingly
complex point sets that ranging from single point to nearest pairs and bigger clusters.\cite{sanchez84}
Any functions defined on the lattice can
be expanded with such a basis. For defective crystals,
the matrix and interstitial sites define such a lattice naturally. If we further introduce a special species of \emph{white atom},
and regard all vacancies as that
occupied by white atoms, then such orthogonal and complete correlation functions $\xi$ can be defined.
In this way, the probability for a configuration to appear becomes
\begin{equation}
  \rho_{q} =\sum_{j}Y_{q,j}\xi_{j},
\end{equation}
and the Gibbs function of Eq.(\ref{eq:gibbs1}) can be rewritten as
\begin{equation}
  G =\sum_{j}v_{j}\xi_{j} -TS_{c},
  \label{eq:G_alloy}
\end{equation}
with the configurational entropy given by $S_{c}=-k_{B}\sum_{q}\left[(\sum_{j}Y_{q,j}\xi_{j})\ln(\sum_{j}Y_{q,j}\xi_{j})\right]$,
and the interaction strength of cluster $v_{j}=\sum_{q}F_{q}Y_{q,j}$.
It is obvious that Eq.(\ref{eq:G_alloy}) has the same form as the theory for alloys.\cite{geng05b,sluiter90,sluiter96} Namely,
both alloying and defects on a lattice can be described by the same unified theoretical framework.
Within this method, the interaction strength of clusters can be evaluated with cluster expansion method
using \emph{ab initio} total energy calculations,\cite{connolly83} and the configurational entropy may be
evaluated by cluster variation method.\cite{kikuchi51}

\subsection{Effective point defect model}

For a point defect on an infinitely large lattice (the dilute limit), the formalism discussed
above naturally leads to the conventional PDM.\cite{matzke87,lidiard66} When defect concentration is finite but interaction between them is
weak and can be ignored safely, the same conclusion holds.
This is because for a configuration containing $y$ non-interacting point defects, the corresponding FGE is just
$y$ times of that of a single one, and then Eq.(\ref{eq:rhoq}) gives rise to $\rho_{q_{y}}=(\rho_{q_{1}})^{y}$,
which is exactly the result of PDM. Alternatively, for non-interacting defects,
the presence of a defect has no influence on others, we therefore can isolate a defect by cutting it
and the associated local lattice out from the matrix, and then extend the surrounding lattice to
an infinite range. This operation keeps the defective behavior.
It maps non-interacting defects with finite concentrations onto a group of systems at the dilute limit, which justifies
the application of PDM.

When interactions between defects are substantial, it is almost impossible to isolate a defect from
others. At a condition that the distribution of defects is homogeneous, however, we can approximate
the interaction by a constant mean-field. In this approximation,
the defects and the associated local lattice environment that are cut out from the matrix
keep the original size, and subject to a field
that takes a role of modeling the interactions with other homogeneously distributed defects
that have been removed. To construct accurately an environmental field of such kind is difficult,
if not impossible. For practical purpose, we may simplify it by using regularly
distributed defects to simulate the field approximately. This sacrifices the rigidity
of the theory, but makes the problem more tractable. What one needs to do now is to periodically repeat the
piece of defective lattice that has been cut out from the matrix along the three dimensional
lattice vectors. In this way the periodical images of the defect take the role of modeling the
homogeneous environment produced by other defects. Just one such configuration
of course can not capture the whole features of the defect interactions.
By averaging over all possible
regular defective distributions, nevertheless, one eventually can reach a converged result.

Generally the free energy of a system in the ensemble can be written as
\begin{equation}
  F_{q}=F_{0} + \sum_{i}A_{i}n_{i}+\frac{1}{2}\sum_{i,j}B_{ij}n_{i}n_{j}+
  \frac{1}{6}\sum_{i,j,k}C_{ijk}n_{i}n_{j}n_{k}+\cdots,
  \label{eq:gibbs2}
\end{equation}
where $n_{i}$ is the concentration of defect $i$ in this system. If only
up to linear terms are kept, then Eqs.(\ref{eq:gibbs1}) and (\ref{eq:rhoq}) reduce
back to the conventional PDM. The terms of higher order describe effective defect interactions, and should
be important for any real materials with high defect density. On the other hand, if we know
the values of parameter $F_{0}$, $A$, $B$, and $C$, the free energy (as a function of defect concentration $n_{i}$) of any defective configuration
can be evaluated from Eq.(\ref{eq:gibbs2}) directly.
To determine these parameters, one can solve Eq.(\ref{eq:gibbs2}) by a least square fit
method using \emph{ab initio} calculated $F_{q}$ of
a set of configurations. Furthermore, since the $\Delta G$ in Eq.(\ref{eq:rhoq}) can be rewritten as
\begin{equation}
\Delta G=\Delta F^{0} +\Delta F(n_{i},n_{j},\cdots)+ \sum_{i}\mu_{m}\Delta N_{i}^{m},
\label{eq:dG}
\end{equation}
where the first term at the right hand side represents the contribution of non-interacting defects,
and the second term arises from defect interactions, Eq.(\ref{eq:rhoq}) then leads to
\begin{equation}
  \rho_{q} = \rho^{0}_{q}(n^{0}_{i})\exp\left(-\frac{\Delta F}{k_{B}T}\right),
  \label{eq:rhoq2}
\end{equation}
where $\rho^{0}$ is the probability predicted by conventional PDM which gives a defect concentration of $n^{0}$ (determined by $\Delta F^{0}$).
For example, if there are $d_{i}$ defects of $i$-th type appearing in the configuration
$q$, where $n_{i} = d_{i}/D_{i}$ and $D_{i}$ is the total available sites for that defect,
PDM gives $\rho^{0}_{q}(n_{i})=\prod_{i} (n_{i})^{d_{i}}$. Because of the structure of $\Delta F$
as shown in Eqs.(\ref{eq:gibbs2}) and (\ref{eq:dG}), $\rho_{q}$ can be factorized into the same form as $\rho_{q}^{0}$.
Namely,
\begin{equation}
  \rho_{q} = \prod_{i} (n_{i})^{d_{i}} =\rho^{0}_{q}(n_{i}).
\end{equation}
We finally get
\begin{equation}
  n_{i} = n_{i}^{0}\exp\left(\frac{-1}{k_{B}TD_{i}}\left(\frac{1}{2}\sum_{j}B_{ij}n_{j}+
  \frac{1}{6}\sum_{j,k}C_{ijk}n_{j}n_{k}+\cdots \right)\right).
\end{equation}
This explicitly demonstrates that the simple formalism of PDM is still valid
even when defect interactions present, as long as
the defect distribution is homogeneous. The effect of interaction is to modify the defect concentrations in a
constant mean-field way (here \emph{constant} means that the interaction
has been averaged over the whole configurational space so that no dependence on the distance between defects
presents explicitly), and the formation energy of a single point defect $\Delta f_{i}$ has to be changed from its dilute limit
value $\Delta f_{i}^{0}$ to
\begin{equation}
  \Delta f_{i}^{0} \rightarrow \Delta f_{i}=\Delta f_{i}^{0} +\frac{1}{2D_{i}}\sum_{j}B_{ij}n_{j}+
  \frac{1}{6D_{i}}\sum_{j,k}C_{ijk}n_{j}n_{k}+\cdots,
  \label{eq:df_int}
\end{equation}
and the defect concentration equations also become
\begin{equation}
  n_{i}^{0}=\exp\left(-\frac{\Delta f_{i}^{0}}{k_{B}T}\right) \rightarrow
  n_{i}=\exp\left(-\frac{\Delta f_{i}}{k_{B}T}\right).
  \label{eq:GPDM}
\end{equation}

\subsection{Behavior near the stoichiometry}

\begin{table}
\caption{\label{tab:coheE} First-principles results for the energy curve of UO$_{2\pm x}$,
where $x$ is the deviation from the stoichiometric composition of uranium dioxide, $N$ is
the total number of atoms in the simulation cell.
$D$ (in eV), $r_{0}$ (in $\mathrm{\AA}$),
and $B_{0}$ (in GPa) are the cohesive energy per atom, the equilibrium lattice parameter of the
effective cubic cell, and the zero pressure bulk modulus, respectively. }
\begin{ruledtabular}
\begin{tabular}{l c c c c c c l} 
  Structure & $x$ & $N$ &Functional & $D$ & $r_{0}$ & $B_{0}$ & Phase \\
\hline
  ${}^{u}C1_{1}$  &$-\frac{2}{5}$&13&LSDA+$U$ &7.541&5.708  &173.60 & CaF$_{2}$\,(AFM)\\
  $C1_{-1}$       &$-\frac{1}{4}$&11&LSDA+$U$ &8.002&5.443  &189.32 & CaF$_{2}$\,(AFM)\\
  $C1_{1}$        &$\frac{1}{4}$&13&LSDA+$U$ &7.937&5.402  &246.63 & CaF$_{2}$\,(AFM)\\
  ${}^{u}C1_{-1}$ &$\frac{2}{3}$&11&LSDA+$U$ &7.616&5.284  &114.59 & CaF$_{2}$\,(AFM)\\
  $C6_{-1}$     &$-\frac{1}{24}$&71&LSDA+$U$ &8.184&5.447  &214.85 & CaF$_{2}$\,(AFM)\\
\end{tabular}
\end{ruledtabular}
\end{table}

In practice, one usually has to employ a finite size cell with periodic boundary conditions to
simulate the defective structures. The formation energy and defect concentrations
thus obtained in most cases do not correspond to the dilute limit.
Making use of the effective PDM generalized in above subsections, we can quantify not only how the
interactions modify defect concentrations, but also the variation of defect formation
energy as a function of defect concentrations, thus provides a viable way to extrapolate
defect energetics to the dilute limit. Taking UO$_{2}$ as a prototype, we will show in this part
how interactions could alter defect behaviors.

\begin{table}
\caption{\label{tab:Ef} Formation energy (in eV) of intrinsic point defects in UO$_{2}$
of Frenkel pairs (O\_FP and U\_FP) and Schottky defect (Sch). $\Delta \overline{f}^{0}$ is the value
approximated with a $2\times2\times2$ supercell, $\Delta f^{0}$ is the dilute limit
value extrapolated using Eq.(\ref{eq:df_int}) up to cubic terms, and $\delta=\Delta \overline{f}^{0}-\Delta f^{0}$.
}
\begin{ruledtabular}
\begin{tabular}{l c c c } 
  Label & O\_FP & U\_FP & Sch \\
\hline
  $\Delta \overline{f}^{0}$   &5.38 &14.34  &10.53 \\
  $\Delta f^{0}$   &4.77 &13.78  &10.21 \\
  $\delta$             &0.61 &0.56   &0.32 \\
\end{tabular}
\end{ruledtabular}
\end{table}

According to Ref.\onlinecite{geng12} and above discussions, pseudo phases in UO$_{2}$ are well defined: at most
interested thermodynamic conditions only one type of defect presents, and all other components
are suppressed completely. This implies that only the diagonal terms in
Eq.(\ref{eq:gibbs2}) make sense. Namely, only interactions between the same kind of defect need to be considered,
which greatly reduces the number of \emph{ab initio} calculations that are required
for extraction of the interaction strengths $B$ and $C$.
Using configurations with different simulation cell size of $2\times2\times2$, $1\times2\times2$,
and $1\times1\times1$ of the cubic fluorite unit, we extracted the defect interaction
strength by solving a set of equations of Eq.(\ref{eq:gibbs2}).
The employed energy curves
for the smallest cell are listed in table \ref{tab:coheE} (in which the defective structures are labeled follwoing
the same rule of Ref.\onlinecite{geng11}), and others are taken from table I
in Ref.\onlinecite{geng11}.
In particular, the results in table \ref{tab:coheE} were calculated using VASP code, with the same
LSDA+\emph{U} setting as in Ref.\onlinecite{geng11}. All structures were fully relaxed at a series
of fixed volumes. Since the supercell size of these structures is relatively small, 36 irreducible
$k$ points were employed to ensure the total energy convergency.

With the effective PDM of Eq.(\ref{eq:GPDM}), it is not necessary to
work on the ensemble average of Eq.(\ref{eq:gibbs1}) any longer. Instead, the problem
changes to ``how the effectively independent defects distribute on the lattice''. For a purpose
of investigating the compression behavior of defects, it is helpful to employ a reference supercell,
and normalize all involved energetics with respect to it.
In doing so, however, the number of defects might no longer be an integer.
A re-scaling procedure is thus required when evaluating
the formation energy of a \emph{single} defect.
Let $\Delta e$ be the defect contribution
in Eq.(\ref{eq:gibbs2}) that is evaluated in the reference cell, \emph{i.e.} $\Delta e=F-F_{0}$, with a defect concentration equals $n$.
The number of unit cells in a supercell which contains \emph{one} and \emph{only one}
of this type of defect is $1/(nN_{d})$, where $N_{d}$ is the number of available sites for
this defect in a unit cell.
Then the energy difference for
creating a defect is $\Delta E = \Delta e/(nN_{d}N_{r})$, where $N_{r}$ is the number of unit cells
making up the reference supercell. In this way, the formation energy of a Frenkel pair for $\mathrm{X}$ species
becomes
\begin{equation}
  \Delta f_{\mathrm{X\_FP}} =\Delta E_{\mathrm{X}_{v}}+\Delta E_{\mathrm{X}_{i}},
\end{equation}
and the Schottky defect formation energy is
\begin{equation}
  \Delta f_{\mathrm{S}} = 2\Delta E_{\mathrm{O}_{v}}+\Delta E_{\mathrm{U}_{v}}+\frac{3}{N}F_{0}.
\end{equation}
Here $N$ is the total number of atoms in a defect-free reference cell. For a $2\times2\times2$ supercell
of fluorite UO$_{2}$, $N=96$ and $N_{r}=8$. Also for a cubic fluorite unit, $N_{d}$ takes 8 for O$_{v}$,
and 4 for O$_{i}$, U$_{v}$, and U$_{i}$, respectively.
These formulations, together with Eq.(\ref{eq:df_int}), allows for extrapolating the
intrinsic defect formation energy to the dilute limit by decrease the defect concentration to
an arbitrarily small value. The obtained
results are summarized in table \ref{tab:Ef}. We can see that a $2\times2\times2$ cell
is not big enough to converge the formation energy to the dilute limit. The deviation from
the extrapolated value is less than 1\,eV. The largest one is oxygen Frenkel pair in which
$\delta$ reaches a value of 0.6\,eV. It is at the same level of the finite-size correction
of charged defects,\cite{crocombette12} where a value of about 0.6\,eV was also obtained for
{O\_FP}. This good agreement demonstrates that our treatment on defect interaction is at least qualitatively correct.

\begin{figure}
  \includegraphics*[width=3.0 in]{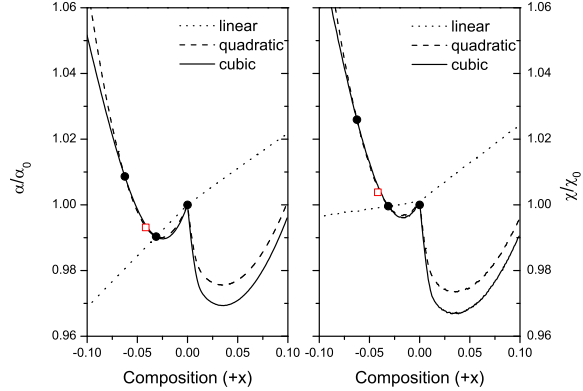}
  \caption{Relative variation of thermal expansivity ($\alpha$) and compressibility ($\chi$)
  with stoichiometry deviation $x$ at different level of approximations. Notice the sharp tips
  at $x=0$ due to defect interactions.
  }
  \label{fig:alpha}
\end{figure}

Inclusion of defect interactions into the PDM makes it possible to study the fine behavior over
the whole stoichiometry. Figure \ref{fig:alpha} shows the relative variation of thermal
expansivity $\alpha$ and compressibility $\chi$ as a function of stoichiometry deviation $x$
at 300\,K and 0\,GPa, in which the solid points denote the exact value of the configurations that
were employed to extract the interaction strength. It is evident that the linear approximation of
Eq.(\ref{eq:gibbs2}) (namely the conventional PDM) fails to reveal the fine behavior of
non-stoichiometric UO$_{2}$. Far from the point that was used to approximate the defect formation energy,
it deviates from the exact value drastically.
On the other hand, both quadratic and cubic approximations
predict a curved variation of physical quantities correctly. It is interesting to note that the sharp tip appearing at the stoichiometry
is very similar to the ``W'' shape anomaly in alloys.\cite{geng05} Nevertheless, the underlying physics
is different. Here it mainly originates from two facts: (a) the curvature
due to defect interactions, and (b) the predominant defect type at the hyper- and hypo-stoichiometry
sides are different, which gives rise to different variation trends of the physical quantities.
From this perspective, measuring the departure of relevant physical quantities
from a linear behavior near the stoichiometry would reveal the strength of defect interactions,
and constrain the application range of the conventional PDM.

Above discussions revealed the power of including defect interaction into statistical mechanics model such as PDM.
Though by comparison with available experimental data one can assess the validity of our approach, due to
the uncertainty of these data as mentioned in Sec.\ref{sec:intr} and \ref{sec:PDM}, a quantitative
validation has to be made with direct \emph{ab initio} calculations. To this end, we made an independent
calculation on a defective configuration of $C6_{-1}$, using a supercell of $1\times2\times3$ of the cubic fluorite unit,
in which one oxygen atom has been removed to create one oxygen vacancy. The obtained energetics are listed in table \ref{tab:coheE},
and the calculated thermal expansivity and compressibility are compared with that of effective
PDM in Fig.\ref{fig:alpha} as the open square points. Note that in this separate calculation we employed
only the experimental observation that oxygen vacancy prevails in UO$_{2-x}$, and did not invoke
any other approximations. Therefore the good agreement between these two results as shown in Fig.\ref{fig:alpha} provides a solid
verification of the validity of our constant mean-field treatment of defect interactions for
homogeneously distributed defects, as well as the effective PDM that based on it.

\section{discussion}
\label{sec:discussion}
\subsection{Charged state}

UO$_{2}$ is a semiconductor with a finite energy gap, it is possible that defects in it are
charged rather than being neutral. Except very few studies,\cite{andersson11,nerikar09a,crocombette11,crocombette12}
most \emph{ab initio} investigations
on defective UO$_{2}$ assumed a neutral simulation cell, as we did here. In principle, such ``neutral''
calculations do not correspond to literally neutral defects, since local transfer of electrons
might lead to a partial charge of the defects. Nevertheless, finite size of the simulation
cell imposes a constraint on the charge redistribution, and thus defect might not reach its full
charge state.
This problem becomes very severe when at the dilute limit or near the stoichiometry,
where defects can be
fully charged only by exchanging electrons with valence/conduction bands, which is a kind of global charge
redistribution.
For large stoichiometry deviation, however,
defects interact with the matrix strongly, leading to deep defect levels which are either at near the gap
middle or hybridization with valence/conduction bands.
In both cases a neutral defect could be expected because it is difficult to \emph{ionize} the defect at low
temperatures in the first case and no charged defect can be supported in the second one.

For oxygen clusters, Crocombette argued that charged state makes them unfavorable at the stoichiometry.\cite{crocombette12}
It is reasonable. Actually, ``neutral'' calculations also indicated that oxygen clusters
have negligible concentrations when near the stoichiometry.\cite{geng08b,geng08c} Putting these information together,
we ascertain that it should have no
defect clustering when $x\approx0$. But this does not mean that oxygen clustering is
negligible at non-stoichiometry. The experimental evidence for such clustering was in fact observed at
large values of $x$,\cite{willis64b,willis78,murray90}
which is compatible with recent neutral calculations that predicted prevailing COT-o clusters
at hyper-stoichiometry region.
On the other hand, \emph{ab initio} electronic structure revealed that the defect levels of COT-o cluster
hybridize strongly with the valence band of the matrix,\cite{geng2010}
which implies that the cluster might be ``neutral'',
or at least these ``neutral'' results should partially reflect some physical reality.
In these considerations, Crocombette's conclusion about charged oxygen defects\cite{crocombette12}
might lose the relevance when far from the stoichiometry.
But overall the charged state of defects in UO$_{2}$
is still an open issue.

In the derivation of the general formalism of defects in Sec.\ref{sec:defect interactions},
we did not consider the charged state. To include this is straightforward. One just needs
to add an additional index to each type of defect to mark its charge state, and include the
chemical potential of free electrons to take the charge contribution at the Fermi level into account.
Eq.(\ref{eq:gibbs1}) then becomes
\begin{equation}
  G=\sum_{q,Q} \left[\rho_{q,Q}\left( F_{q,Q} +Q\epsilon_{f}\right) +k_{B}T\rho_{q,Q}\ln\left(\rho_{q,Q}\right)\right],
  \label{eq:g_charge}
\end{equation}
where $Q$ is the total charge of the system, and $\epsilon_{f}$ the Fermi level. From
this expression, the effective PDM which includes both defect interactions and variable charge
state of defects can be derived easily.

\subsection{Detection of PPT}
\begin{figure}
  \includegraphics*[width=3.5 in]{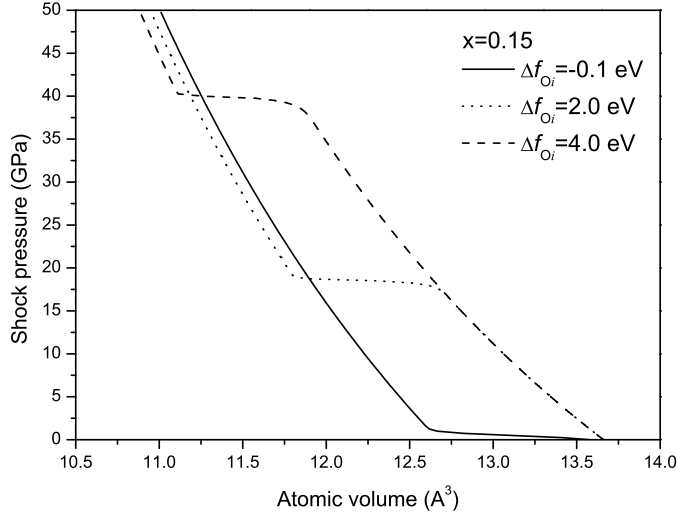}
  \caption{Dependence of pseudo-transition pressure of from uranium vacancy to oxygen interstitial along a shock Hugoniot
  on the change of formation energy of oxygen interstitial. The stoichiometry deviation is $x=0.15$.
    }
  \label{fig:PV}
\end{figure}

Although PPT and the corresponding boundaries can be employed to constrain/extract the dilute limit
of defect formation energy---an important quantity for understanding the stoichiometric behavior,
to measure these boundaries is not a trivial work. At big stoichiometric deviation,
the volume change at PPT is prominent, thus it can be
detected by measuring the quasi-discontinuous jumps in the EOS of defects.\cite{geng12}
Figure \ref{fig:PV} plots a compression curve of UO$_{2+0.15}$ along a Hugoniot shocked
from 300\,K and 0\,GPa. The volume collapse due to PPT from U$_{v}$ to O$_{i}$ (here we ignored
oxygen clustering) is evident and therefore detectable. On the other hand,
this pseudo transition pressure depends sensitively on
the formation energy of $\mathrm{O}_{i}$: it spans over
a wide range of 40\,GPa when there is a change in $\Delta f_{\mathrm{O}_{i}}$ about 4.0\,eV.
This property
guarantees a good precision for the constraints on defect energetics.

At small value of $x$, however, the volume jump would be too weak to be perceptible. This is
usually the case when $|x|<0.02$.\cite{geng12} In these cases, we cannot locate the PPT via measuring
thermodynamic or mechanical quantities. However, since PPT changes the predominant defect
species and thus the position of the defect level within the energy gap, transport properties are
also modified. We therefore can detect the occurring of a PPT by measuring the sudden changes in
electrical conductivity (or optical properties).\cite{stratton87} This method
has high sensitivity so that allows us to access the vicinity of the stoichiometry.

\subsection{Instantaneous response}

\begin{figure}
  \includegraphics*[width=3.0 in]{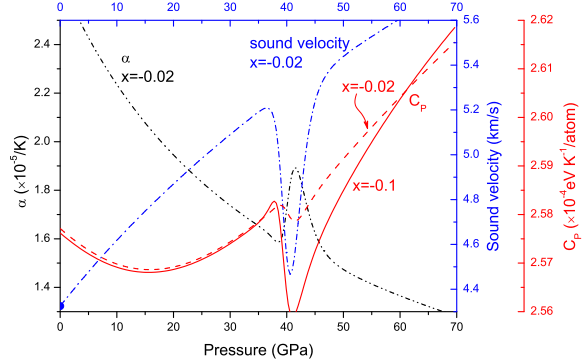}
  \caption{(Color online) Anomalies in the thermal expansivity ($\alpha$),
  isothermal bulk sound velocity, and specific heat at constant pressure ($C_{P}$) along a
  Hugoniot that shocked from 500\,K and 0\,GPa in non-stoichiometric UO$_{2}$ when defect concentrations
  have an instantaneous response to thermal fluctuations. The solid circle (at the bottom left corner) marks the experimental bulk
  sound velocity measured at ambient conditions
  in perfect UO$_{2}$.
  }
  \label{fig:CpP}
\end{figure}

In above discussions and also in Ref.\onlinecite{geng12}, we froze the defect concentrations
when evaluating the thermodynamic quantities. It is a theoretical requirement for the first
derivatives of the Gibbs function, such as volume and entropy. But for higher order derivatives of the Gibbs
function, it has no reason to do so because they are also defined by thermodynamic relations.
In practice, however, a justification for this operation can  be made.
This is because for a classical crystal, the change of defect species can proceed via
only atomic diffusion, which is a very slow process, and thus
no defect can respond to rapid thermal fluctuations.

Then an interesting question arises, that is what about if defects can instantaneously respond to any
disturbances?
Simple analysis shows it might be fantastic. At first the magnitude of anomalies
due to PPT would be amplified greatly,
thus ease the difficulty in PPT detections.
Figure \ref{fig:CpP} demonstrates this effect
on the thermal expansivity $\alpha$, bulk sound velocity, and specific heat at constant pressure
$C_{P}$. The influence can be fully comprehended by comparing with the Fig.2 in Ref.\onlinecite{geng12}, where
defect concentrations were fixed when evaluating these quantities.
Secondly, the compressibility would diverge at zero temperature. Considering the
relationship between sound velocity and the compressibility, this implies
a vanishing sound velocity (and the bulk modulus)
in the vicinity of a PPT at low temperatures if defects have instantaneous response.
On the phonon spectrum, it would manifest as an abnormal softening in
acoustic branches at long-wave length (\emph{i.e.}, $\Gamma$ point in the reciprocal space).
This observation is tantalizing. But can it be true? We cannot answer it yet.
Classical atomic diffusion of course can not lead to a rapid response. But what if for a quantum
crystal? In a quantum world particles are described by a wave-function. If the
wavefunctions of all defects are in a coherent state, then an instantaneous and simultaneous change of the defect species
might be possible.\cite{pushkarov91} In this mechanism the requirement for
a large scale atomic diffusion has been removed, and a single change in defecton state
is enough for a quantum pseudo transition to occur. Nevertheless, much more theoretical works are necessary
in order to pin down this possibility definitely.

\section{conclusion}

A general formalism for
the thermodynamics of defects in a crystal was derived
based on the statistics of a grand canonical ensemble on a lattice.
By introducing idle white atoms for vacancy and
extending the sublattice of interstitial sites, this formalism has
the same form as the lattice
theory for alloys and compounds---a reflection of the unified
physics underlying these seemingly different systems.
With an approximation of
constant mean-field, this
theory reduces to an effective point defect model in which defect interactions are included
by an auxiliary field, whereas each individual defects are treated as independent. In this way,
we mapped a \emph{many-body} defect system onto a \emph{single} defect system by coupling it with an effective
external field. If ignoring this field, the conventional PDM is recovered.
This generalization greatly expands the applicability of the simple PDM.
In order to explore the full content in this theory, we also studied possible reentrant
PPT and multi-defect coexistence with virtual systems.

Using PDM, we investigated the possibility of constraining defect energetics by measuring
pseudo phase boundaries. By calculating the possible PPT between interested defects, we showed that
the experimental estimates available in literature, as well as variety theoretical assessments,
on defective energetics of UO$_{2}$, are not fully consistent. On the other hand, the
range of energetics constrained by the PPTs overlaps with these estimates largely, therefore has a potential to
reduce the inconsistency in these data.
By including defect interactions into PDM, we demonstrated that the information obtained at finite stoichiometry deviation
can be extrapolated to the dilute limit.
Finally, we investigated the fine
behavior of thermal expansivity and compressibility in the vicinity of the stoichiometry of defective UO$_{2}$,
and some relevant issues of charged defects, detection of PPT, and possible instantaneous
response of defectons in a quantum crystal are briefly discussed, in which we highlighted
the detection of PPT by measuring electrical conductivity when near the stoichiometry,
and the complexity arising from possible charged state of defects. Through these investigations,
we clearly demonstrated that it is valuable to explore the whole non-stoichiometric range
in order to acquire a comprehensive understanding about a defective material thoroughly.

\begin{acknowledgments}
The authors acknowledge financial support from the National Natural Science Foundation of China under
Grant No.11274281, and the research Project 2012A0101001 supported by CAEP.
\end{acknowledgments}


\end{document}